# Comparing System Dynamics and Agent-Based Simulation for Tumour Growth and its Interactions with Effector Cells

Grazziela P. Figueredo, Uwe Aickelin

IMA Research Group, Computer Science School, Nottingham University, Wollaton Road, Nottingham, NG8 1BB UK

gzf@cs.nott.ac.uk, uxa@cs.nott.ac.uk

**Keywords:** system dynamics simulation, agent-based simulation, immune system simulation, comparison of system dynamics and agent-based simulation.

## Abstract

There is little research concerning comparisons and combination of System Dynamics Simulation (SDS) and Agent Based Simulation (ABS). ABS is a paradigm used in many levels of abstraction, including those levels covered by SDS. We believe that the establishment of frameworks for the choice between these two simulation approaches would contribute to the simulation research. Hence, our work aims for the establishment of directions for the choice between SDS and ABS approaches for immune system-related problems. Previously, we compared the use of ABS and SDS for modelling agents' behaviour in an environment with no movement or interactions between these agents. We concluded that for these types of agents it is preferable to use SDS, as it takes up less computational resources and produces the same results as those obtained by the ABS model. In order to move this research forward, our next research question is: if we introduce interactions between these agents will SDS still be the most appropriate paradigm to be used? To answer this question for immune system simulation problems, we will use, as case studies, models involving interactions between tumour cells and immune effector cells. Experiments show that there are cases where SDS and ABS can not be used interchangeably, and therefore, their comparison is not straightforward.

## 1. INTRODUCTION

The current scenario in the simulation field presents paradigms that allow us to build simulation models for various domains. Some of the important simulation approaches are System Dynamics Simulation (SDS), Agent-Based Simulation (ABS), Discrete Event Simulation (DES) and Dynamic Systems (DS)[3]. New research also combines these methods and defines frameworks for the usage of each paradigm. There is already work comparing SDS/DES, DES/ABS as well as their combinations. However, there is few research on the comparison and combination of SDS and ABS [8]. Hence, our research aims at establishing a framework for the development of simulations involving the choice between SDS and ABS approaches and their combination for immune system-related problems .

ABS is a paradigm used in many levels of abstraction, including those levels covered by SDS. As there is this intersection, some range of simulation problems can be solved by either SDS or ABS. In previous work [1], we compared the use of ABS and SDS for modelling static agents' behaviour in an immune system ageing problem. By static we mean that there is no movement or interactions between the agents. We concluded that for these types of agents, it is preferable to use SDS instead of ABS. When contrasting the results of both simulation approaches, we see that SDS is less complex and takes up less computational resources, producing the same results as those obtained by the ABS model.

To advance this study, our next inquiring is: Once we have established that SDS is more suitable for static agents than ABS, if we introduce interactions between these agents will SDS still be the most appropriate paradigm to be used?

To answer this question for immune system simulation problems, we use models as case studies, which include interactions between tumour cells and immune effector cells. Effector cells are responsible for eliminating tumour cells in the organism. Our goal is to determine which scenarios for immune system simulations inside the SDS/ABS intersection would benefit from SDS resources and those that are more suitable for ABS.

For SDS we need to establish mathematical equations that determine the flows. Therefore we use the models reviewed in [4]. The authors of this study explored existing spatially homogeneous mechanistic mathematical models describing the interactions between a malignant tumor and the immune system. They begin with the simplest (single equation) models for tumor growth and proceed to consider greater immunological detail (and correspondingly more equations) in steps. For our simulation, we intend to build SDS and ABS for two of the most important mathematical models described in [4]. We intend to use the mathematical model also as basis for the ABS. The idea is to check if the results would be similar and if we can use SDS and ABS for our case studies interchangeably.

This work is organized as follows. In Section 2, there is the related work relevant to our studies. Section 3 presents the mathematical models used for our simulations, as well as the simulations we carried out and their results. Finally, in

Section 4 we draw the conclusions and present future work.

## 2. RELATED WORK

The theoretical work presented by [3] compares ABS and SD conceptually, and discusses the potential synergy between them in order to solve problems of teaching decision-making processes. The authors explore the conceptual frameworks for ABS and SD to model group learning. They show the conceptual differences between these two paradigms and propose their use in a complementary way. They outstand the lack of knowledge in multi-paradigm simulation involving SD and ABS.

In [6] and [8], the authors present a cross-study of SD and ABS. They define their features and characteristics and contrast the two methods. Moreover, they also present ideas of how to integrate both approaches. As a continuation of this work, in [7] they present an approach to integrate the SD and ABS techniques for supply chain management problems. They present some preliminary results, which were not the same as the SD simulation alone. Therefore, they propose new tests as future work. In their case study, they were not able to reach the same results with both simulations.

In [9], the authors show the application of both SD and ABS to simulate non-equilibrium ligand-receptor dynamics over a broad range of concentrations. They concluded that both approaches are powerful tools and are also complementary. In their case study, they did not indicate a preferred paradigm, although intuitively SD is an obvious choice when studying systems at a high level of aggregation and abstraction. On the other hand, SD is not capable of simulating receptors and molecules and their individual interactions, which can be done with ABS.

Rahmandad and Sterman [5] compare the dynamics of a stochastic ABS model with those of the analogous deterministic compartment differential equation model for contagious disease spread. The authors convert the ABS into a differential equation model and examine the impact of individual heterogeneity and different network topologies. The deterministic model yields a single trajectory for each parameter set, while stochastic models yield a distribution of outcomes. Moreover, the differential equation model and ABS dynamics differ for several metrics relevant to public health. The response of the models to policies can also differ when the base case behaviour is similar. Under some conditions, however, the differences in means are small, compared to variability caused by stochastic events, parameter uncertainty and model boundary.

In our previous work [1], we compared SDS and ABS for a naive T cell output model. In that study, we concluded that for that case study SDS is more suitable. We had a scenario where the agents had no interactions and SDS and ABS produced similar outputs. Therefore, we decided that, in this case, it is preferable to choose the SDS, as it takes up less computational resources. In order to continue the investigation, which compares these two simulation approaches, we will add interactions between agents and compare the results. The description of the problem, as well as the mathematical equations can be seen in the next section.

## 3. MATHEMATICAL MODELS

In this section, we present mathematical models used as basis for our simulations. The simplest ones involve only one equation and defines mathematical rules for tumour growth (Section 3.1.). The second mathematical model addresses the interactions between tumour cells and immune effector cells. This is shown in Section 3.2..

### 3.1. One-Equation Models: Tumour Growth

In this section, we present the simplest mathematical models. They have only one equation that describes how tumours grow. There are three models of tumour growth from [4] considered in this study: the logistic model, the von Bertalanffy model and the Gompertz model.

According to [4], the most general equation describing the dynamics of tumor growth can be written as:

$$\frac{dT}{dt} = T f(T), \qquad (1)$$

where:

- $T$ is the tumour cell population at time $t$,
- $T(0) > 0$,
- $f(T)$ specifies the density dependence in proliferation and death of the tumour cells. The density dependence factor can be written as:

$$f(T) = p(T) - d(T), \qquad (2)$$

where:

- $p(T)$ defines tumour cells proliferation,
- $d(T)$ define tumour cells death.

The expressions for $p(T)$ and $d(T)$ are generally defined by power laws:

$$p(T) = aT, \qquad (3)$$

$$d(T) = bT, \qquad (4)$$

For our experiments, we defined the values for    and using the three well established models:

**Logistic Model:** $\alpha = 0$ and $\beta = 1$ ($a, b > 0$ and $b < a$ for growth),

**von Bertalanffy Model:** $\alpha = \frac{1}{3}$ and $\beta = 0$ ($a, b > 0$ and $b < a$ for growth),

**Gompertz Model:** $p(T) = a$ and $d(x) = b\ln(T)$ ($a, b > 0$ and $e^b > a$ for growth).

### 3.1.1. SDS for the One-Equation Model

We have implemented the one-equation models using SDS. Figure 1 shows the stock and flow diagram used for modelling the mathematical equations. From the SDS perspective, the amount of tumour cells is a stock that will be modified by proliferation and death flows.

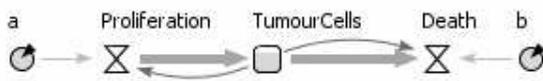

**Figure 1.** SDS diagram for the general one-equation mathematical model.

### 3.1.2. ABS for the One-Equation Model

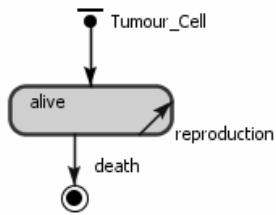

**Figure 2.** One equation agent.

We have also implemented the one-equation model using ABS. In this case we have a tumour cell agent that will reproduce and die, as it can be seen in Figure 2.

### 3.1.3. Experiments

We carried out two experiments to compare the SDS and ABS simulation outputs. For the first experiment, we have established a variable $c$, which represents the ratio between $a$ and $b$. The purpose of $c$ is to observe the impact of $a$ and $b$ on the tumour growth curve. Therefore, we set $c$ as 5, 2.5, 1.7 and 1.25.

In the second experiment, we defined $a = 1.636$ and $b \in \{0.002, 0.005\}$. These values were determined in [2] and they will be used for our next simulation set of experiments using a two-equation model.

As the outcomes for ABS are stochastic, we ran each simulation for 50 times and the mean simulation output is presented. For both simulations, we defined the initial values for tumour cells equal to 1.

### 3.1.4. Results

For the first experiment, as is shown in Figure 3 (middle and bottom), the outputs for both simulation approaches are similar for the von Bertalanffy and Gompertz models. For the logistics model, on the other hand, the results of the ABS did not match the SDS ones (Figure 3 top). This can be explained by the stochastic and individual behaviour of the agents in the ABS model.

In the experiments, there are few agents on the logistics model and most of them die before they reproduce. In this ABS model, the reproduction rate is given by $p(T) = a$, because $alpha = 0$ for the logistics model. In the case where $a = 1$ and $b = 0.2$, for example, when the number of agents becomes bigger than 5, $d(T)$ gets greater than $p(T)$. As for ABS the death rate is defined for the agents individually, all the newborn tumour cells after 5 agents in the system will have death rates bigger than reproduction rates. Hence, the agents population disappears over the first days. This difference on the SDS and ABS results lead us to conclude that there are cases using static agents where SDS and ABS outputs will not be the same.

Nevertheless, if we increase the difference between the parameters $a$ and $b$, as is shown in Figure 4, we avoid the premature death of the tumour agents and therefore the ABS and SDS results become similar again. From what we have found on the literature, the Logistics model is one of the most used for average tumours whereas the von Bertalanffy and Gompertz models are used for more aggressive tumours. Therefore, as we increase in the difference between $a$ and $b$, the Gompertz and von Bertalanffy models growth will present a considerable increase on the proliferation of tumour cells, as illustrates Figure 5.

Therefore, we could not carry out experiment two using ABS for Gompertz and von Bertalanffy models. From the SDS results, we can see that the number of tumour cells bypasses the magnitude of $10^{64}$ in the Gompertz model (Figure 3). To run the same experiment with ABS we would need more computational resources and it would take up more processing time. In our case, as each agent takes up around 1 megabyte of memory, we would need a memory capacity of $10^{64}$ megabytes. Therefore, in this case it is preferable to run the simulation using SDS, even though such big number of tumour cells also seems to be unrealistic in tumour biology.

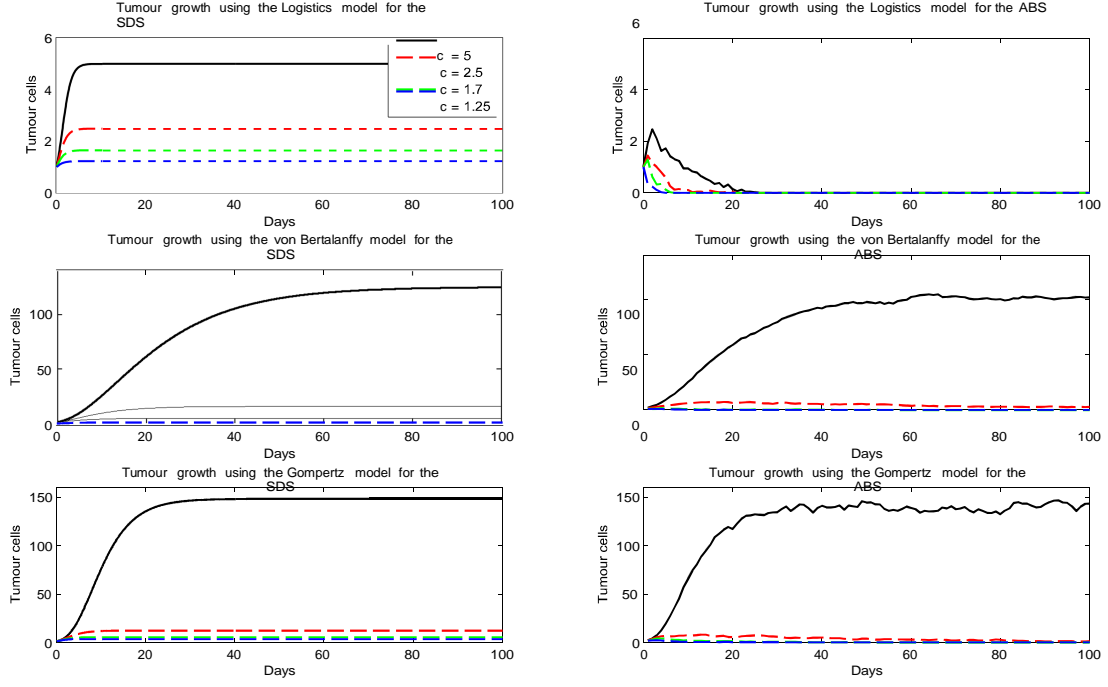

**Figure 3.** Results for the one-equation model using SDS and ABS.

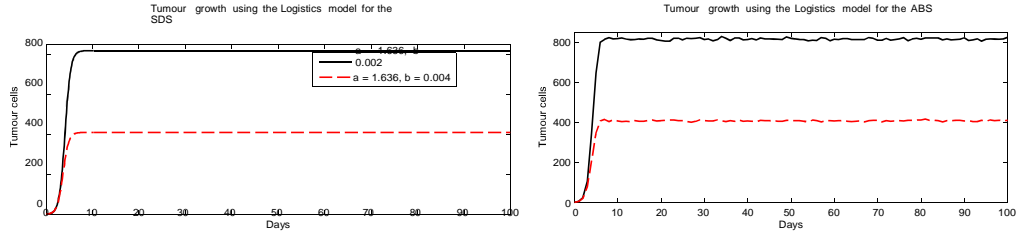

**Figure 4.** Results for the second experiment of one-equation Logistics model using SDS and ABS.

### 3.2. Two-Equation Models: Interaction Between Tumour Cells and Generic Effector Cells

To add complexity to our model, we are going to consider tumour cells growth together with their interactions with general immune effector cells. In this stage we are not yet considering specific types of immune cells. Effector cells are responsible for killing the tumour cells. They proliferate and die per apoptosis, which is a programmed cellular death. In the models, cancer treatment is also considered.

The interactions between tumour cells and immune effector cells can be defined by the equations:

$$\frac{dT}{dt} = T f(T) - d_T(T, E) \quad (5)$$

$$\frac{dE}{dt} = p_E(T, E) - d_E(T, E) - a_E(E) + \ (T), \quad (6)$$

where:

- $T$ is the number of tumour cells,
- $E$ is the number of effector cells,
- $f(T)$ is the growth of tumour cells,
- $d_T(T, E)$ is the number of tumour cells killed by effector cells,
- $p_E(T, E)$ is the proliferation of effector cells,
- $a_E(E)$ is the death (apoptosis) of effector cells.

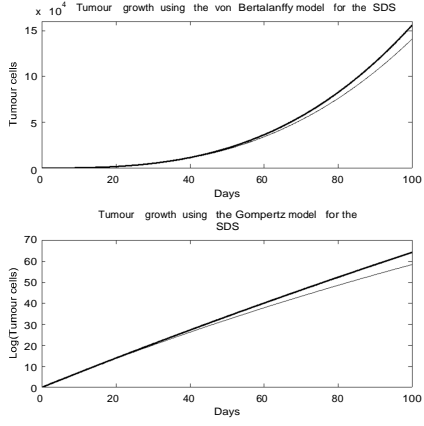

**Figure 5.** Results for the second experiment of one-equation Gompertz and Von Bertalanffy models using SDS.

- $(T)$ is the treatment or influx of cells.

For our models, we used the Kuznetsov model [2]:

$$f(T) = a(1 - bT), \quad (7)$$

$$d_T(T, E) = nTE, \quad (8)$$

$$p_E(E, T) = \frac{pTE}{g + T}, \quad (9)$$

$$d_E(E, T) = mTE, \quad (10)$$

$$a_E(E) = dE, \quad (11)$$

$$(t) = s. \quad (12)$$

As it can be seen, the Logistic model was adopted for tumour growth. It seems to be the most common model of tumour growth used in the mathematical models involving cancer and the immune system. The values for the parameters on the equations can be seen in Table 1. We got these values from [4]. In the first three scenarios we consider cancer treatment. The fourth case does not consider any treatment.

### 3.2.1. SDS for the Two-Equation Model

We have converted the mathematical model into a SDS model. Figure 6 shows the stock and flow diagram we have defined.

We consider two stocks, the tumour cells and the immune effector cells. The tumour cell stock is changed by proliferation and natural death (as defined in the one-equation model in Section 2.1); and death caused by the immune effector cells. The immune cells stock is changed by death, apoptosis, proliferation and injection of new cells as treatment. The number of tumour cells in the organism also stimulates the proliferation of immune cells.

| Scenario | b | d | s |
|---|---|---|---|
| 1 | 0.002 | 0.1908 | 0.318 |
| 2 | 0.004 | 2 | 0.318 |
| 3 | 0.002 | 0.3743 | 0.1181 |
| 4 | 0.002 | 0.3743 | 0 |

**Table 1.** Simulation parameters for different scenarios. For the other parameters, the values are the same in all experiments, i.e, $a = 1.636$, $g = 20.19$, $m = 0.00311$, $n = 1$ and $p = 1.131$.

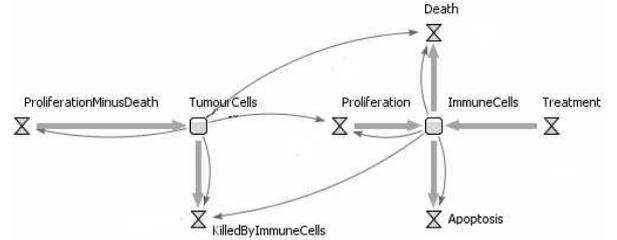

**Figure 6.** SDS diagram for the two-equation mathematical model.

### 3.2.2. ABS for the Two-Equation Model

For the ABS model, we defined two agents that will interact with each other: the tumour cell agent and the effector cell agent. Figure 7 shows the state charts for the effector cells and the tumour cells. The state chart for the effector cells (left hand side of Figure 7) has two states. Either the cell is alive and able to reproduce or is dead. Effector cells can die by natural means or by damage. The tumour cells state chart also has two states. Tumour cells can reproduce, die with age or die killed by effector cells. The rates defined in the transitions are the same of the mathematical model.

### 3.2.3. Experiments

As we mentioned before, we carried out four experiments for the two-equation model. The parameter variation for the experiments is shown in Table 1. In the four scenarios, it is considered differences in the death rate of tumour cells (defined by the parameter $b$), effector cells apoptosis rate (defined by the parameter $d$) and treatment (parameter $s$).

Similar to the one-equation model, we ran the simulation

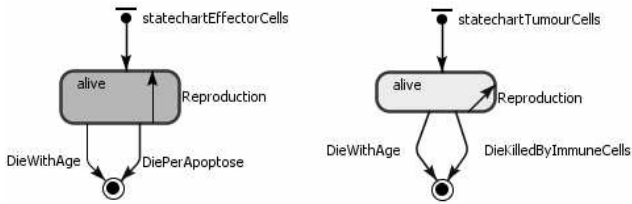

**Figure 7.** ABS diagram for the two-equation mathematical model. On the left hand side we have the state chart for the effector cell agents and on the right hand side we have the state chart for the tumour cell agents.

for the ABS 50 times and display the mean values as the results.

### 3.2.4. Results

The results comparing SDS and ABS for the four experiments can be seen in Figure 8 (first and second columns).

For the first scenario, the behaviour of the tumour cells is very similar for the SDS and ABS results. However, when we ran the Wilcoxon test for the tumour cells outcome, it rejected the similarity hypothesis for both outcomes, as shown in Table 2. This might be due to the fact that the tumour cells for the SDS model decrease in an asymptotic way towards to zero, while ABS behaviour is discrete and hence, it reaches zero. The number of effector cells for both simulations follows the same pattern, although the numbers are not the same. The variances in the ABS curve were expected due to its stochastic character.

The results for the second scenario seem to be similar for effector cells, although the Wilcoxon test rejects this hypothesis. The results for the tumour cells are not the same.

For scenarios 3 and 4, the results are completely different for both simulation approaches. Moreover, when we look at the tumour cells curve, the differences are even more evident. In scenario 3 using SDS, tumour cells decrease as effector cells increase, following a predator-prey trend curve. This output is what would be expected by the mathematical model. On the other hand, for the ABS, the number of effector cells decreases until a value close to zero while the tumour cells numbers vary in time differently from the SDS results. The predator-prey behaviour is not observed in this experiment.

In scenario 4, although effector cells seem to decay in a similar trend for both approaches, the results for tumour cells are completely different. In the SDS simulation, the numbers of tumour cells reach a value close to zero after twenty days and then increases again. For the ABS simulation, on the other hand, tumour cells reach zero and never increase again. It happens because SDS deals with continuous stocks and ABS has a discrete number of agents.

The results in these experiments show that there are simulation cases where SDS and ABS derived from the same mathematical model do not have the same output. Therefore, it is not possible to compare which approach would be more suitable for these cases.

One alternative would be the development of an ABS solution, which is not based on the rates defined in the mathematical model. However, it seems that for each output (or parameter change on the mathematical model), there should be a different ABS implementation.

For example, if we determine that the number of tumour cells should always be bigger than zero in the ABS, for the second and fourth scenarios the outputs become closer to the SDS, as shown in Figure 8. On the other hand, this constraint also changes the first scenario results, which seemed satisfactory before. For scenario 4, although the outcome with this fix do not seem very similar on the beginning of the simulation, the steady state has closer values.

To achieve similar results for scenario 3, we also had to determine that the number of effector cells should be bigger than zero. The results are shown in line 4 of Figure 8. For scenario 3 the fix did not work perfectly. It seems that only the steady state of the simulation has closer results. We also tried to randomly add some effector cells over time, but the results did not looked similar as well.

## 4. CONCLUSIONS

The current scenario in the simulation field presents paradigms that allow us to build simulation models for various domains. New research compares simulation methods and defines frameworks for the usage of each paradigm. However, there is few research on the comparison and combination of SDS and ABS [8]. We aim at contributing to this area by studying immune system simulation problems. Therefore, in this study, we used case studies which include interactions between tumour cells and immune effector cells. Our goal was to determine which scenarios for immune system simulations inside the SDS/ABS intersection would benefit from SDS resources and those that are more suitable for ABS.

The models we used were reviewed in [4]. We began with the simplest (single equation) models for tumor growth and proceed to consider tow-equation models involving effector and tumour cells. We used mathematical models as basis for both ABS and SDS. The idea was to check if the results would be similar and if we can use SDS and ABS for our case studies interchangeably.

We carried out two experiments to compare the outputs for the one-equation model. For the first experiment, the outputs for both simulation approaches are similar for two cases. There was an example, though, where the results of the ABS did not match the SDS's. This is explained by the stochastic and individual behaviour of the agents in the ABS model.

To add complexity to our tests, we considered tumour cells growth together with their interactions with general immune effector cells. We defined four scenarios and, for only one of them, the results were similar using the mathematical model. The differences in the output are due to the fact that effector cells numbers change continuously on the SDS, while for the ABS, they change in a discrete pattern.

| Implementation | Cells | Scenario | | | | | | | |
| --- | --- | --- | --- | --- | --- | --- | --- | --- | --- |
| | | 1 | | 2 | | 3 | | 4 | |
| | | p | h | p | h | p | h | p | h |
| *ABS* | *Tumour* | 0 | 1 | 0 | 1 | 0 | 1 | 0 | 1 |
| | *Effector* | 0.0028 | 1 | 0.0325 | 1 | 0 | 1 | 0 | 1 |
| *ABS - Fix 1* | *Tumour* | 0 | 1 | 0.0103 | 1 | 0 | 1 | 0.8595 | 0 |
| | *Effector* | 0 | 1 | 0.4441 | 1 | 0 | 1 | 0 | 1 |
| *ABS - Fix 2* | *Tumour* | 0 | 1 | 0 | 1 | 0 | 1 | 0 | 1 |
| | *Effector* | 0 | 1 | 0 | 1 | 0 | 1 | 0 | 1 |

**Table 2.** Wilcoxon test for tumour cells and effector cells. Comparing the results between SDS and ABS.

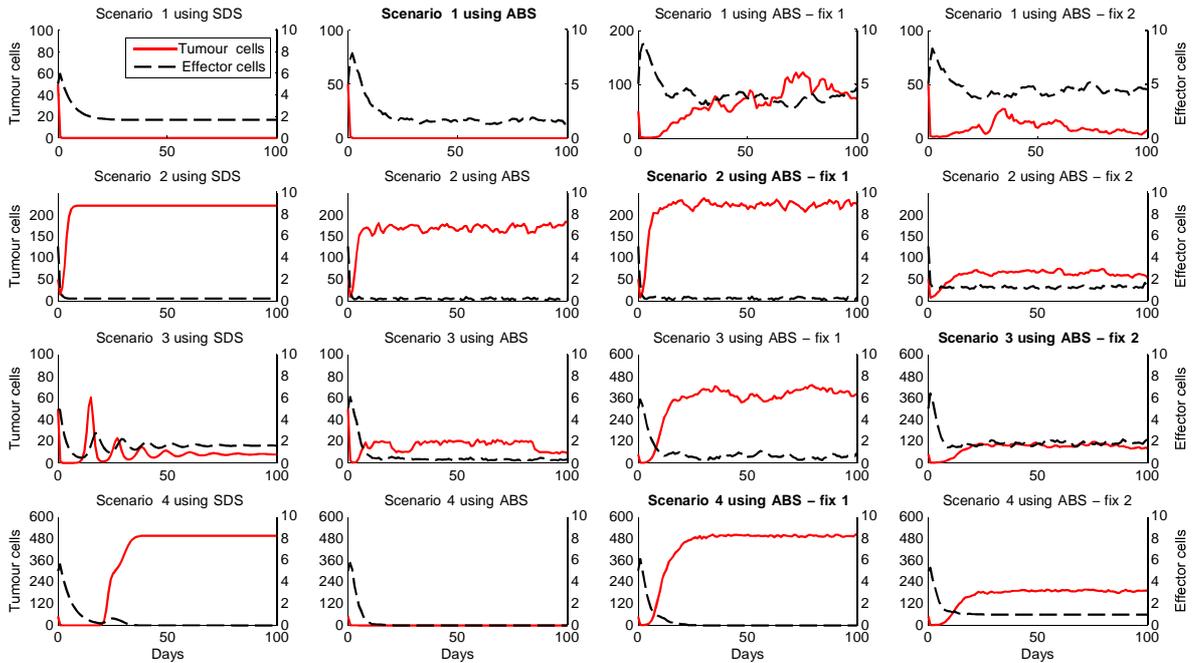

**Figure 8.** Results for the two-equation mathematical model using SDS and ABS. On the first columns we have the SDS results for the four scenarios. The second column has the ABS results. The third and fourth columns shows the fixes we tried to make the outcomes similar. Fix 1 adds the constraint *tumour cells* > 0. Fix 2 adds the constraint *tumour cells* > 0 and *effector cells* > 0. For each graph we have the results for tumour cells (continuous line) and effector cells (dotted lines). The scale for the tumour cells results is on the left y axis of each graph while the scale for effector cells is in the right y axis.

The results in these experiments show that there are simulation cases where SDS and ABS derived from the same mathematical model do not have the same output. Therefore, it is not possible to compare which approach would be more suitable for these cases.

One alternative would be the development of an ABS solution, which is not based on the rates defined in the mathematical model. However, it seems that for each output (or parameter change on the mathematical model), there should be a different ABS implementation.

As future work we intend to work with models with three an four equations and compare the results of SDS and ABS without using the mathematical equation as baseline.